\documentclass[11pt]{article}
\usepackage{amsmath,amsfonts,amssymb,graphicx,epsfig,pdflscape,multirow,theorem,tikz,tikz-cd}
\usepackage{bm,cite}
\usepackage{pstricks}
\usepackage{simplewick}
\usetikzlibrary{decorations.markings}
\tikzset{middlearrow/.style={
        decoration={markings,
            mark= at position 0.5 with {\arrow{#1}} ,
        },
        postaction={decorate}
    }
}
\numberwithin{equation}{section}
\graphicspath{{./eps/}}
\usepackage{braket}
\usepackage{mathtools}

\theorembodyfont{\itshape} 
\theoremheaderfont{\scshape}
\theoremstyle{plain}

\numberwithin{equation}{section}

\newcommand{\nc}{\newcommand}
\nc{\bib}{\bibitem}
\nc{\be}{\begin{equation}}
\nc{\ee}{\end{equation}}
\nc{\id}{\mathbb{I}}
\nc{\eb}{\bm{\epsilon}}
\nc{\ob}{\bm{\omega}}
\nc{\inb}{\bm{\infty}}
\nc{\gr}{\mathrm{gr}}
\nc{\g}{\mathfrak{g}}
\nc{\gh}{\widehat{\mathfrak{g}}}
\nc{\Ac}{\mathcal{A}}
\nc{\Bc}{\mathcal{B}}
\nc{\Ic}{\mathcal{I}}
\nc{\Oc}{\mathcal{O}}
\nc{\Qc}{\mathcal{Q}}
\nc{\Vir}{\mathfrak{Vir}}
\nc{\pa}{\partial}
\nc{\eps}{\epsilon}
\nc{\ii}{\mathrm{i}}
\nc{\z}{\mbox{{\gray {\tiny 0}}}}
\nc{\gast}{\!\ast}
\nc{\s}{\;\!\!}
\nc{\La}{\Lambda}

\begin{document}

\topmargin -15mm
\oddsidemargin 05mm

%%%%%%%%%%%%%%%%%%%%%
%
% Title page
%
%%%%%%%%%%%%%%%%%%%%%

\title{\mbox{}\vspace{0in}
\bf 
\huge
Multi-graded Galilean conformal algebras
\\[-.8cm]
}
\date{}

\maketitle

\begin{center}
{\vspace{-5mm}\LARGE Eric Ragoucy$^*$,\, J{\o}rgen Rasmussen$^{**}$,\, Christopher Raymond$^{**}$}
\\[.4cm]
{\em ${}^{*}$Laboratoire d'Annecy-le-Vieux de Physique Th\'eorique (LAPTH)}\\
{\em Universit\'e Grenoble Alpes, CNRS, F-74000 Annecy, France}
\\[.4cm]
{\em ${}^{**}$School of Mathematics and Physics, University of Queensland}\\
{\em St Lucia, Brisbane, Queensland 4072, Australia}
\\[.4cm] 
{\tt ragoucy\,@\,lapth.cnrs.fr}
\qquad
{\tt j.rasmussen\,@\,uq.edu.au}
\qquad
{\tt christopher.raymond\,@\,uqconnect.edu.au}
\end{center}

%%%%%%%%%%%%%%%%%%%%%
%
% Abstract
%
%%%%%%%%%%%%%%%%%%%%%

\vspace{0.5cm}
\begin{abstract}
Galilean conformal algebras can be constructed by contracting a finite number of conformal 
algebras, and enjoy truncated $\mathbb{Z}$-graded structures. 
Here, we present a generalisation of the Galilean contraction procedure, giving rise to
Galilean conformal algebras with truncated $\mathbb{Z}^{\otimes\sigma}$-gradings, $\sigma\in\mathbb{N}$.
Detailed examples of these multi-graded Galilean algebras are provided, including
extensions of the Galilean Virasoro and affine Kac-Moody algebras. 
We also derive the associated Sugawara constructions and discuss how these examples relate to multivariable 
extensions of Takiff algebras.
We likewise apply our generalised contraction prescription to tensor products of $W_3$ algebras and obtain new
families of higher-order Galilean $W_3$ algebras. 
\end{abstract}

\vfill
\begin{flushleft}
LAPTH-004/20
\end{flushleft}

%\newpage
%\tableofcontents

\newpage
%%%%%%%%%%%%%%%%%%%%%%%%%%
\section{Introduction}
\label{Sec:Introduction}
%%%%%%%%%%%%%%%%%%%%%%%%%%

The Galilean Virasoro algebra appears in studies of asymptotically flat three-dimensional spacetimes, 
see \cite{BO14} and references therein, and can be constructed \cite{BC07,BG09,HR10,BGMM10,BF12} as 
a contraction of a pair of Virasoro algebras. 
Similarly, the Galilean $W_3$ algebra \cite{ABFGR13,GMPT13,ChrisThesis,CGOR16,RR17} follows
by contracting a pair of $W_3$ algebras, while more general Galilean conformal algebras with extended 
symmetries have been constructed in \cite{GRR15,ChrisThesis,RR17} and are known as Galilean $W$-algebras.
Non-relativistic systems with (typically non-affine) conformal 
symmetry were previously studied in \cite{Hag72,Nie72,Hen94,NORM97,LSZ06,NS07,DH09}.

Following ideas put forward in \cite{RR17}, higher-order Galilean contractions were developed in \cite{RR19}, 
generalising the contraction procedure from pairs of symmetry algebras (or equivalently vertex algebras)
to any finite number of symmetry algebras (or vertex algebras).
In the case of Virasoro or affine Kac-Moody, the usual second-order Galilean algebras have been 
found \cite{RR19} to be isomorphic to the Takiff algebras \cite{Tak71} considered in \cite{HSSU12,BR13}, 
while the higher-order counterparts provide $N$th-order generalisations 
(where $N$ is the number of inputted symmetry algebras $\Ac$).
These higher-order Galilean algebras thus enjoy a truncated $\mathbb{Z}$-grading whose truncation 
is determined by the order $N$ of the contraction.

Here, we modify the higher-order contraction procedure to let it depend on a factorisation of $N$, where 
${\bf N}=N_1,\ldots,N_\sigma$ is a finite sequence of positive integers such that $N=N_1\ldots N_\sigma$.
We thus organise the $N$-fold tensor product $\Ac^{\otimes N}$ in terms of $N_\ell$-fold tensor-product factors,
\be
 \Ac^{\otimes N}=\big(\Ac^{\otimes N_1}\big)\otimes\ldots\otimes\big(\Ac^{\otimes N_\sigma}\big),
\ee
and apply the higher-order contraction prescription of \cite{RR19} to the factors `simultaneously'.
We find that the ensuing Galilean algebra, 
$\Ac_G^{\bf N}$, is $\mathbb{Z}^{\otimes\sigma}$-graded, truncated according to the sequence $N_1,\ldots,N_\sigma$.
Because of this graded structure, we refer to the generalised contraction as {\em multi-graded contraction}.
We also observe that the contractions are independent of the ordering of the factors in the factorisation of $N$.
In case the mode algebra underlying $\Ac$ is a Lie algebra, we find that $\Ac_G^{\bf N}$ is isomorphic to a 
multivariable generalisation of the Takiff algebras discussed in \cite{RR19}, 
with the number of variables given by the length $\sigma$
of the contraction sequence ${\bf N}$.

In Section \ref{Sec:Contraction}, we outline the multi-graded contraction procedure and illustrate it by 
working out the corresponding Galilean Virasoro and affine Kac-Moody algebras.
We also discuss the ensuing grading structures and relate the corresponding Galilean algebras 
to a multivariable generalisation of the Takiff algebras.
In Section \ref{Sec:GenSug}, we construct a Sugawara operator for each of the multi-graded Galilean 
Kac-Moody algebras; its central charge is given by the product of the contraction
order $N$ and the dimension of the underlying Lie algebra. We also show that the Sugawara construction 
commutes with the contraction procedure.
In Section \ref{Sec:GW3}, we apply multi-graded contractions to the $W_3$ algebra and thereby obtain a new
class of Galilean $W_3$ algebras.
Section \ref{Sec:Discussion} contains some concluding remarks.

%%%%%%%%%%%%%%%%%%%%%%%%%%
\section{Contraction procedure}
\label{Sec:Contraction}
%%%%%%%%%%%%%%%%%%%%%%%%%%

We find it advantageous to describe the multi-graded contractions and ensuing algebras in the language of 
operator-product algebras (OPAs), and refer to \cite{Thi95,RR17} for details on the structure of an OPA.
We say that an OPA is of Lie type if the corresponding mode algebra is a Lie algebra, as is the case for the 
Virasoro and Kac-Moody algebras.
Throughout, $\id$ denotes the identity field and $\Delta_A$ the conformal weight of the scaling field $A$.

%%%%%%%%%%%%%%%%%%%%%%%%%%
\subsection{Star relations in OPAs}
\label{Sec:Star}
%%%%%%%%%%%%%%%%%%%%%%%%%%

For the space of quasi-primary fields in the OPA $\Ac$, we let $\Bc_{\Ac}$ denote a basis consisting of 
quasi-primary fields only. Only keeping the non-singular terms, the operator-product expansion 
of $A,B\in\Bc_{\Ac}$ can then be expressed as
\be
 A(z)B(w) \sim \sum_{Q\in\Bc_{\Ac}} C_{A,B}^Q\left( \sum_{n=0}^{\Delta_A+\Delta_B-\Delta_Q}
  \frac{\beta^{\Delta_Q;n}_{\Delta_A,\Delta_B}\partial^nQ(w)}{(z-w)^{\Delta_A+\Delta_B-\Delta_Q- n}} \right),
\label{AB}
\ee
with structure constants $C^{Q}_{A,B}\in\mathbb{C}$ and 
\be
\beta^{\Delta_{Q};n}_{\Delta_{A},\Delta_{B}} = \frac{(\Delta_{A} - \Delta_{B} + \Delta_{Q})_{n}}{n!(2\Delta_{Q})_{n}},
\qquad (x)_{n} = \prod_{j=0}^{n-1} (x+j).
\ee
Convenient for our purposes, the essential part of the operator-product expansion (\ref{AB}) is synthesised in 
the so-called star relation
\be
 A \ast B \simeq \sum_{Q\in\Bc_{\Ac}} C_{A,B}^Q\{Q\},
\label{ABQ}
\ee
where $\{Q\}$ represents the sum over $n$ displayed in (\ref{AB}).

For example, the Virasoro algebra $\Vir$ of central charge $c$ is of Lie type and generated by $T$, 
with star relation
\be
 T \ast T \simeq \tfrac{c}{2}\{\id\} + 2 \{T\}.
\label{Vir}
\ee
Likewise, the nontrivial star relations in an affine Kac-Moody algebra $\gh$ are given by
\be
 J^a\ast J^b\simeq\kappa^{ab}k\,\{\id\}+{f^{ab}}_c\{J^c\},
\label{KM}
\ee
where ${f^{ab}}_c\in\mathbb{C}$ are structure constants, $k\in\mathbb{C}$ the level 
and $\kappa$ the Killing form of the underlying finite-dimensional complex Lie algebra $\g$.
As is customary, we do not display summations over repeated group indices, 
here the summation over $c\in\{1,\ldots,\dim\g\}$.
We note that $\gh$ is of Lie type.

%%%%%%%%%%%%%%%%%%%%%%%%%%
\subsection{Higher-order contractions}
\label{Sec:Higher}
%%%%%%%%%%%%%%%%%%%%%%%%%%

Higher-order Galilean contractions were developed in \cite{RR17}. Here, we recast them in a notation suitable
for their multi-graded generalisation introduced in Section \ref{Sec:Gen}. Thus, for $N\in\mathbb{N}$, let
\be
 \Ac^{\otimes N}=\bigotimes_{i=0}^{N-1}\Ac_{(i)},
\ee
where $\Ac_{(0)},\ldots,\Ac_{(N-1)}$ are copies of the same OPA $\Ac$, up to the values of their central 
parameters (such as central charges or levels). In effect, we are viewing the central parameters of $\Ac$
as independent indeterminants. We then write
\be
 {\bf A}_{*}=\begin{pmatrix} A_{(0)}\\ \vdots\\ A_{(N-1)}\end{pmatrix},\qquad
 {\bf c}_{*}=\begin{pmatrix} c_{(0)}\\ \vdots\\ c_{(N-1)}\end{pmatrix},
\ee
where $A_{(i)}$ (respectively $c_{(i)}$) denotes the field $A\in\Ac_{(i)}$ (respectively a central parameter of $\Ac_{(i)}$).
For $\eps\in\mathbb{C}$, we also let
\be
 U_N(\eps,\omega)=D_N(\eps)U_N(\omega),\qquad
 D_N(\eps)=\mathrm{diag}(\eps^0,\eps^1,\ldots,\eps^{N-1})
\ee
and
\be
 U_N(\omega)=\begin{pmatrix}\omega^{ij}\end{pmatrix}_{0\leq i,j\leq N-1}
 =\begin{pmatrix}
 \omega^0&\omega^0&\cdots&\omega^0\\[.1cm]
 \omega^0&\omega^1&\cdots&\omega^{N-1}\\[.1cm]
 \vdots&\vdots&\ddots&\vdots\\[.1cm]
  \omega^0&\omega^{N-1}&\cdots&\omega^{(N-1)^2}
 \end{pmatrix},
\ee
where $\omega$ is the principal $N$th root of unity,
\be
 \omega=e^{2\pi\ii/N}.
\ee
It follows that (for $\eps\neq0$)
\be
 D_N^{-1}(\eps)=D_N(\eps^{-1}),\qquad U_N^{-1}(\omega)=\tfrac{1}{N}U_N(\omega^{-1}).
\ee
Thus, with
\be
 {\bf A}_{\eps}=\begin{pmatrix} A_{0,\eps}\\ \vdots\\ A_{N-1,\eps}\end{pmatrix}=U_N(\eps,\omega){\bf A}_{*},\qquad
 {\bf c}_{\eps}=\begin{pmatrix} c_{0,\eps}\\ \vdots\\ c_{N-1,\eps}\end{pmatrix}=U_N(\eps,\omega){\bf c}_{*},
\ee
the map
\be
 \Ac^{\otimes N}\to\Ac^{\otimes N},\qquad {\bf A}_{*}\mapsto{\bf A}_{\eps},\qquad{\bf c}_{*}\mapsto{\bf c}_{\eps},
\label{AA}
\ee
is invertible for $\eps\neq0$. For $\eps=0$, on the other hand, the map is singular (unless $N=1$).
If a well-defined OPA arises in the limit $\eps\to0$, where
\be
 {\bf A}_{\eps}\to{\bf A}=\begin{pmatrix} A_{0}\\ \vdots\\ A_{N-1}\end{pmatrix},\qquad
 {\bf c}_{\eps}\to{\bf c}=\begin{pmatrix} c_{0}\\ \vdots\\ c_{N-1}\end{pmatrix},
\ee
the ensuing algebra is known \cite{RR17} as the {\em $N$th-order Galilean OPA} $\Ac^N_G$.
Note that $\Ac_G^1\cong\Ac$.
For small $N$, the Galilean Virasoro algebras $\Vir_G^N$ also appeared in \cite{CCRSR17}.

%%%%%%%%%%%%%%%%%%%%%%%%%%
\subsection{Generalised higher-order contractions}
\label{Sec:Gen}
%%%%%%%%%%%%%%%%%%%%%%%%%%

Fix $\sigma\in\mathbb{N}$. As in Section \ref{Sec:Introduction}, we denote complex-number sequences of length 
$\sigma$ by ${\bf S}=S_1,\ldots,S_\sigma$ etc, with ${\bf 0}=0,\ldots,0$ the zero sequence. 
Linear combinations are readily formed
\be
 \alpha\,{\bf i}+\beta\,{\bf j}=\alpha i_1+\beta j_1,\ldots,\alpha i_\sigma+\beta j_\sigma,\qquad 
  \alpha,\beta\in\mathbb{C},
\ee
and two sequences can be compared as
\be
 {\bf i}\leq{\bf j}\quad\mathrm{if}\quad i_1\leq j_1,\ldots,i_\sigma\leq j_\sigma;\qquad
  {\bf i}<{\bf j}\quad\mathrm{if}\quad i_1<j_1,\ldots,i_\sigma<j_\sigma.
\ee
If every element of ${\bf S}$ is nonzero, we let ${\bf S}^{-1}$ denote the sequence 
$S_1^{-1},\ldots,S_\sigma^{-1}$.

The set
\be
 I_{\bf N}=\{{\bf i}\in\mathbb{Z}^{\otimes\sigma}\,|\,{\bf 0}\leq{\bf i}<{\bf N}\}
\label{i}
\ee
of integer sequences bounded strictly by ${\bf N}$ admits the {\em canonical order} where
${\bf i}$ appears before ${\bf j}$ if and only if for each $m$ such that 
${i_m}>{j_m}$ there exists $\ell<m$ such that ${i_\ell}<{j_\ell}$.
This corresponds to the usual ordering of basis vectors for the tensor product space
$V=V_1\otimes\ldots\otimes V_\sigma$, where, for each $\ell\in\{1,\ldots,\sigma\}$, $V_\ell$
is an $N_\ell$-dimensional vector space with ordered basis $\{e_1^\ell,\ldots,e_{N_\ell}^\ell\}$.
That is, in the $N$-vector formed by the components $\{v_{\bf i}\,|\,{\bf 0}\leq{\bf i}<{\bf N}\}$ in the 
decomposition
\be
 {\bf v}=\sum_{{\bf 0}\leq{\bf i}<{\bf N}}v_{\bf i}e_{\bf i}\in V,
\label{v}
\ee
the components are ordered according to the canonical ordering of the multi-indices ${\bf i}$.
For example, for ${\bf N}=2,3$, the components are ordered as
\be
 v_{0,0},\,v_{0,1},\,v_{0,2},\,v_{1,0},\,v_{1,1},\,v_{1,2}.
\ee

Using the same ordering prescription, we now form the $N$-dimensional vectors
\be
 {\bf A}_{*}=\begin{pmatrix} A_{({\bf i})}\end{pmatrix}_{{\bf 0}\leq{\bf i}<{\bf N}},\qquad
 {\bf c}_{*}=\begin{pmatrix} c_{({\bf i})}\end{pmatrix}_{{\bf 0}\leq{\bf i}<{\bf N}},
\label{Astar}
\ee
where $A_{({\bf i})}$ (respectively $c_{({\bf i})}$) denotes the field $A\in\Ac_{({\bf i})}$ 
(respectively a central parameter of $\Ac_{({\bf i})}$).
With
\be
 \ob=\omega_1,\ldots,\omega_\sigma,\qquad \omega_\ell=e^{2\pi\ii/N_\ell},\qquad \ell=1,\ldots,\sigma,
\ee
and for
\be
 \eb=\eps_1,\ldots,\eps_\sigma,\qquad \eps_\ell\in\mathbb{C},\qquad \ell=1,\ldots,\sigma,
\ee
we also introduce
\be
 U_{\bf N}(\eb,\ob)=U_{N_1}(\eps_1,\omega_1)\otimes\ldots\otimes U_{N_\sigma}(\eps_\sigma,\omega_\sigma)
 =D_{\bf N}(\eb)U_{\bf N}(\ob),\qquad
\ee
where
\be
 D_{\bf N}(\eb)=D_{N_1}(\eps_1)\otimes\ldots\otimes D_{N_\sigma}(\eps_\sigma),\qquad
 U_{\bf N}(\ob)=U_{N_1}(\omega_1)\otimes\ldots\otimes U_{N_\sigma}(\omega_\sigma).
\ee
The map
\be
 \Ac^{\otimes N}\to\Ac^{\otimes N},\qquad {\bf A}_{*}\mapsto{\bf A}_{\eps}
  =\begin{pmatrix} A_{{\bf i},\eps}\end{pmatrix}_{{\bf 0}\leq{\bf i}<{\bf N}}=U_{\bf N}(\eb,\ob){\bf A}_{*},\qquad
 {\bf c}_{*}\mapsto{\bf c}_{\eps}=\begin{pmatrix} c_{{\bf i},\eps}\end{pmatrix}_{{\bf 0}\leq{\bf i}<{\bf N}}
  =U_{\bf N}(\eb,\ob){\bf c}_{*},
\label{AAA}
\ee
is invertible if and only if $\eps_1,\ldots,\eps_\sigma\neq0$, in which case
\be
 U^{-1}_{\bf N}(\eb,\ob)=\tfrac{1}{N}U_{\bf N}(\ob^{-1})D_{\bf N}(\eb^{-1}).
\ee
If a well-defined ($N$th-order Galilean) OPA arises in the limit $\eb\to{\bf 0}$, where
\be
 {\bf A}_{\eps}\to{\bf A},\qquad
 {\bf c}_{\eps}\to{\bf c},
\ee
we denote it by $\Ac^{{\bf N}}_G$.

For $U_{\bf N}(\eb,\ob)$ invertible, using 
\be
 \sum_{n=0}^{N_\ell-1}\omega_\ell^{nk}=N_\ell\delta_{k,0\;\mathrm{mod}\;N_\ell},\qquad \ell=1,\ldots,\sigma,
\ee
we see that, for ${\bf 0}\leq{\bf i},{\bf j},{\bf m}<{\bf N}$,
\be
 \sum_{{\bf 0}\leq{\bf k}<{\bf N}}U_{\bf N}(\eb,\ob)_{\bf i\,k}\,U_{\bf N}(\eb,\ob)_{\bf j\,k}\,
  U_{\bf N}(\eb^{-1},\ob^{-1})_{\bf k\,m}=N\delta_{{\bf m},{\bf i}+{\bf j}}.
\label{UUU}
\ee
In Section \ref{Sec:Multi}, we use this result to determine the structure of $\Ac_G^{\bf N}$ for $\Ac$ of Lie type.

%%%%%%%%%%%%%%%%%%%%%%%%%%
\subsection{Multi-grading}
\label{Sec:Multi}
%%%%%%%%%%%%%%%%%%%%%%%%%%

Still treating central parameters as indeterminants, we assign the following grades to the generators and 
parameters of the Galilean algebras:
\be
 \gr:\Ac_G^{\bf N}\to I_{\bf N},\qquad A_{\bf i}\mapsto{\bf i},\qquad c_{\bf i}\mapsto{\bf i}.
\label{gr}
\ee
The action of $\gr$ is then extended linearly and to rational functions of the central parameters and 
normal-ordered products and derivatives of the fields, with $\mathrm{gr}(\pa)=0$, so that, for instance,
\be
 \gr\Big(c_{0,0}\pa B_{3,1}-35\,\frac{c_{1,1}c_{2,1}+4c_{3,2}}{c_{1,4}}\,(A_{0,1}B_{1,2})\!\Big)=3,1.
\ee
We say the algebra is {\em multi-graded} if the grading is compatible with the product structure of 
operator-product expansions, in the sense that
\be
 \gr(A_{\bf i}\ast B_{\bf j})={\bf i}+{\bf j}.
\label{grAB}
\ee
A priori, it is not guaranteed that all terms appearing in the decomposition of $A_{\bf i}\ast B_{\bf j}$ have a 
well-defined grade, let alone the same grade. However, as we will argue, all Galilean algebras of the type 
$\Ac_G^{\bf N}$ are, in fact, multi-graded. Moreover, the grading is finitely truncated by ${\bf N}$ in the sense 
that $A_{\bf i}\ast B_{\bf j}=0$ unless ${\bf i}+{\bf j}<{\bf N}$.

Let $A,B\in\Bc_{\Ac}$ and consider the star relation (\ref{ABQ}). 
If $\Ac$ is of Lie type, then the only structure constants $C_{A B}^Q$ that can depend on central parameters 
have $Q=\id$, as in (\ref{Vir}) and (\ref{KM}). To indicate this, we write
\be
 A \ast B \simeq C_{A,B}^\id(c)\{\id\}+\sum_{Q\in\Bc_{\Ac}\setminus\{\id\}} C_{A,B}^Q\{Q\},
\label{ABI}
\ee
where $C_{A,B}^\id(c)$ is linear in $c$,
\be
 C_{A,B}^\id(c)=f_{A,B}\,c,\qquad f_{A,B}\in\mathbb{C},
\ee
while $C_{A,B}^Q$ is independent of $c$ for all $Q\in\Bc_{\Ac}\!\setminus\!\{\id\}$.
From (\ref{UUU}), it follows that
\be
 A_{{\bf i},\eps}\ast B_{{\bf j},\eps}=f_{A,B}\,c_{{\bf i}+{\bf j},\eps}\{\id\}
  +\sum_{Q\in\Bc_{\Ac}\setminus\{\id\}} C_{A B}^Q\{Q_{{\bf i}+{\bf j},\eps}\},
\ee
so, in the Galilean algebra $\Ac_G^{\bf N}$,
\be
 A_{\bf i}\ast B_{\bf j}=f_{A,B}\,c_{{\bf i}+{\bf j}}\{\id\}+\sum_{Q\in\Bc_{\Ac}\setminus\{\id\}} 
  C_{A B}^Q\{Q_{{\bf i}+{\bf j}}\}
  =\sum_{Q\in\Bc_\Ac}C_{A B}^Q\{Q_{{\bf i}+{\bf j}}\}.
\ee
Thus, $\Ac_G^{\bf N}$ is multi-graded if $\Ac$ is of Lie type.

The nonlinearity of an OPA that is {\em not} of Lie type obscures the question of its grading structure, 
as witnessed in sections \ref{Sec:GenSug} and \ref{Sec:GW3}. However, as already indicated, 
all the Galilean algebras we have analysed are nevertheless multi-graded in the sense outlined above.

\paragraph{Virasoro algebras:}
The multi-graded Galilean Virasoro algebra $\Vir_G^{\bf N}$ is generated by the fields 
$\{T_{\bf i}\,|\,{\bf 0}\leq{\bf i}<{\bf N}\}$ and has central parameters $\{c_{\bf i}\,|\,{\bf 0}\leq{\bf i}<{\bf N}\}$, 
with star relations given by
\be
T_{\bf i}\ast T_{\bf j}\simeq
\begin{dcases}
\tfrac{c_{{\bf i}+{\bf j}}}{2}\{\id\}+2\{T_{{\bf i}+{\bf j}}\}, \ &{\bf i}+{\bf j}<{\bf N}, \\[.15cm] 
 0, \ &\mathrm{otherwise}.
 \end{dcases}
\label{TiTj}
\ee
Note that $T_{\bf 0}$ generates a subalgebra isomorphic to $\Vir$ with central charge $c_{\bf 0}$,
and that, for every ${\bf i}$, $T_{\bf i}$ is quasi-primary with respect to $T_{\bf 0}$.

\paragraph{Affine Kac-Moody algebras:}
The multi-graded Galilean Kac-Moody algebra $\gh_G^{\,\bf N}$ is generated by 
$\{J_{\bf i}^a\,|\,a=1,\ldots,\dim\g;\,{\bf 0}\leq{\bf i}<{\bf N}\}$, with nontrivial star relations
\be
 J^a_{\bf i}\ast J^b_{\bf j}\simeq\kappa^{ab}k_{{\bf i}+{\bf j}}\{\id\}
  +{f^{ab}}_c\{J^c_{{\bf i}+{\bf j}}\},\qquad  {\bf i}+{\bf j}<{\bf N}.
\ee
Note that $\{J_{\bf 0}^a\,|\,a=1,\ldots,\dim\g\}$ generates a subalgebra isomorphic to $\gh$ at level $k_{\bf 0}$.

%%%%%%%%%%%%%%%%%%%%%%%%%%
\subsection{Permutation invariance}
\label{Sec:Permutation}
%%%%%%%%%%%%%%%%%%%%%%%%%%

In all the Galilean algebras we have analysed, we observe that
\be
 A_{\bf i}\ast B_{\bf j}\simeq A_{\bf i'}\ast B_{\bf j'}\qquad\mathrm{if}\qquad{\bf i}+{\bf j}={\bf i'}+{\bf j'}.
\label{ijij}
\ee
Together with the grading property, this implies that all inequivalent decompositions of star relations arise as 
$A_{\bf 0}\ast B_{\bf j}$ for some $A,B\in\Ac$ and ${\bf 0}\leq{\bf j}<{\bf N}$.
It also implies that the multi-graded contraction procedure is independent of the {\em ordering} of the elements
in the contraction sequence. That is, 
\be
 \Ac_G^{\bf N}\cong\Ac_G^{\pi({\bf N})},\qquad \pi({\bf N})=N_{\pi_1},\ldots,N_{\pi_\sigma},
\ee
where $\pi=\pi_1,\ldots,\pi_\sigma$ is a permutation of the integers $1,\ldots,\sigma$.
Moreover, as the tensorial structure of the contraction process ensures that
\be
 \Ac_G^{N_1,N_2}\cong(\Ac_G^{N_1})_G^{N_2},
\ee
we see that
\be
 (\ldots((\Ac_G^{N_1})_G^{N_2})\ldots)_G^{N_\sigma}\cong\Ac_G^{N_1,\ldots,N_\sigma}
 \cong\Ac_G^{N_{\pi_1},\ldots,N_{\pi_\sigma}}\cong(\ldots((\Ac_G^{N_{\pi_1}})_G^{N_{\pi_2}})\ldots)_G^{N_{\pi_\sigma}}.
\ee

%%%%%%%%%%%%%%%%%%%%%%%%%%
\subsection{Multivariable Takiff algebras}
\label{Sec:Takiff}
%%%%%%%%%%%%%%%%%%%%%%%%%%

For some $R^\ell_{\ell'}\in\mathbb{R}$, $\ell,\ell'=1,\ldots,\sigma$, we let ${\bf N}\to\inb$ denote the limit where
\be
 N_\ell\to\infty,\qquad
 \frac{N_\ell}{N_{\ell'}}\to R^\ell_{\ell'},\qquad \ell,\ell'=1,\ldots,\sigma.
\ee
In this limit, the algebra $\gh_G^{\,\bf N}$ becomes $\gh_G^{\,\inb}$ generated by 
$\{J_{\bf i}^a\,|\,a=1,\ldots,\dim\g;\,{\bf i}\geq{\bf 0}\}$,
with nontrivial star relations
\be
 J^a_{\bf i}\ast J^b_{\bf j}\simeq\kappa^{ab}k_{{\bf i}+{\bf j}}\{\id\}
  +{f^{ab}}_c\{J^c_{{\bf i}+{\bf j}}\}.
\ee
This $\mathbb{Z}^{\otimes\sigma}$-graded algebra is seen to be isomorphic to a multivariable polynomial ring,
\be
 \gh_G^{\,\inb}\,\cong\,\gh\otimes\mathbb{C}[t_1,\ldots,t_\sigma],
\ee
and we likewise recognise the isomorphism
\be
 \gh_G^{\,{\bf N}}\cong\gh\otimes\mathbb{C}[t_1,\ldots,t_\sigma]/
  \langle t_1^{N_1},\ldots,t_\sigma^{N_\sigma}\rangle.
\ee
This extends to multiple variables the Takiff algebras considered in \cite{RR19}, themselves extensions to 
general order $N$ of the second-order (one-variable) Takiff algebras considered in \cite{HSSU12,BR13}. 
We similarly have
\be
 \Vir_G^{\inb}\,\cong\,\Vir\otimes\mathbb{C}[t_1,\ldots,t_\sigma],\qquad 
  \Vir_G^{\bf N}\cong\Vir\otimes\mathbb{C}[t_1,\ldots,t_\sigma]/\langle t_1^{N_1},\ldots,t_\sigma^{N_\sigma}\rangle.
\ee

Further generalisations of the Galilean and Takiff algebras are obtained as follows. 
Let $s=\{s_1,\ldots,s_\rho\}$ denote a subset of $\{1,\ldots,\sigma\}$ and ${\bf N}\to\inb_s$ the limit
where
\be
 N_{s_1},\ldots N_{s_\rho}\to\infty,\qquad
 \frac{N_{s_i}}{N_{s_j}}\to R^{s_i}_{s_j},\qquad s_i,s_j\in s.
\ee
In this limit, the Galilean algebra $\Ac_G^{\bf N}$ becomes $\Ac_G^{\inb_s}$, where, for example,
 \begin{align}
 \gh_G^{\,{\inb_s}}&\cong\gh\otimes\mathbb{C}[t_1,\ldots,t_\sigma]/
  \langle t_{s_1}^{N_{s_1}},\ldots,t_{s_\rho}^{N_{s_\rho}}\rangle,
 \\[.15cm]
   \Vir_G^{\inb_s}&\cong\Vir\otimes\mathbb{C}[t_1,\ldots,t_\sigma]/
     \langle t_{s_1}^{N_{s_1}},\ldots,t_{s_\rho}^{N_{s_\rho}}\rangle.
\end{align}

%%%%%%%%%%%%%%%%%%%%%%%%%%
\section{Generalised Sugawara construction}
\label{Sec:GenSug}
%%%%%%%%%%%%%%%%%%%%%%%%%%

The objective here is to construct a Sugawara operator for each Galilean affine Kac-Moody algebra 
$\gh^{\,{\bf N}}_G$ and to show that this process commutes with the Galilean contraction procedure, 
thereby establishing the commutativity of the diagram
\begin{center}
\begin{tikzcd}[column sep=large,row sep = large]
 \gh^{\,\otimes {\bf N}} \arrow[r, "\mathrm{Sug}^{\otimes {\bf N}}"] \arrow[d,two heads, "\mathrm{Gal}"]
&\Vir^{\otimes {\bf N}} \arrow[d,two heads, "\mathrm{Gal}"] \\
\gh^{\,{\bf N}}_G \arrow[r, "\mathrm{Gal\; Sug}"]
& \Vir^{\bf N}_G
\end{tikzcd}
\end{center}
The lower branch is analysed in Section \ref{Sec:ConSug}; the upper one in Section \ref{Sec:SugCon}.

%%%%%%%%%%%%%%%%%%%%%%%%%%
\subsection{Galilean Sugawara construction}
\label{Sec:ConSug}
%%%%%%%%%%%%%%%%%%%%%%%%%%

For the Sugawara generators of $\Vir_G^{\bf N}$, we make the ansatz
\be
 T_{\bf i}=\sum_{{\bf 0}\leq{\bf r},{\bf s}<{\bf N}}
  \lambda^{{\bf r};{\bf s}}_{\bf i}\kappa_{ab}(J_{\bf r}^aJ_{\bf s}^b),\qquad 
 {\bf 0}\leq{\bf i}<{\bf N},
\label{Ti}
\ee
where $\kappa_{ab}$ are elements of the inverse Killing form on $\g$, and our goal is to determine
the coefficients $\lambda^{{\bf r},{\bf s}}_{\bf i}$ such that
\be
 T_{\bf i}\ast J_{\bf j}^a\simeq\begin{dcases} \{J_{{\bf i}+{\bf j}}^a\},\ &{\bf 0}\leq{\bf i}+{\bf j}<{\bf N},\\[.15cm] 
 0,\ &\mathrm{otherwise}.\end{dcases}
\label{TJ}
\ee
To this end, we compute the operator-product expansion
\begin{align}
 J_{\bf j}^a(z)T_{\bf i}(w)&\sim\frac{1}{{(z-w)^2}}\sum_{{\bf 0}\leq{\bf r},{\bf s}<{\bf N}}\lambda^{{\bf r};{\bf s}}_{\bf i}
  \big[k_{{\bf j}+{\bf r}}J_{\bf s}^a(w)+k_{{\bf j}+{\bf s}}J_{\bf r}^a(w)+2h^\vee J_{{\bf j}+{\bf r}+{\bf s}}^a(w)\big]
 \nonumber\\[.15cm]
 &+\frac{1}{z-w}\sum_{{\bf 0}\leq{\bf r},{\bf s}<{\bf N}}
  \lambda^{{\bf r};{\bf s}}_{\bf i}\kappa_{bc}
  \big[{f^{ab}}_d(J_{{\bf j}+{\bf r}}^dJ_{\bf s}^c)(w)+{f^{ac}}_d(J_{\bf r}^bJ_{{\bf j}+{\bf s}}^d)(w)\big],
\label{JT}
\end{align}
where the dual Coxeter number $h^\vee$ of $\g$ has arisen through
\be
 \kappa_{bc}{f^{ab}}_d{f^{dc}}_e=2h^\vee\delta_e^a.
\ee
To satisfy (\ref{TJ}), the sum multiplying the single pole in (\ref{JT}) must be zero while the sum multiplying 
the double pole must equal $J_{{\bf i}+{\bf j}}^a(w)$. The single-pole constraint implies that
\be
 \lambda^{{\bf r};{\bf s}}_{\bf i}=\begin{dcases}\lambda_{\bf i}^{{\bf n};{\bf N}-{\bf 1}},\ & 
  {\bf r}+{\bf s}={\bf N}-{\bf 1}+{\bf n}\quad ({\bf 0}\leq{\bf n}<{\bf N}),\\[.15cm]
  0,\ & \mathrm{otherwise},
  \end{dcases}
\ee
where ${\bf 1}=1,\ldots,1$. For each ${\bf 0}\leq{\bf i}<{\bf N}$, this fixes all but the $N$ coefficients 
$\lambda_{\bf i}^{{\bf n};{\bf N}-{\bf 1}}$ labelled by ${\bf 0}\leq{\bf n}<{\bf N}$. The double-pole constraint 
then requires that
\be
 2\sum_{{\bf j}\leq{\bf m}<{\bf N}}\sum_{{\bf 0}\leq{\bf n}\leq{\bf m}-{\bf j}}\lambda_{\bf i}^{{\bf n};{\bf N}-{\bf 1}}
  k_{\,{\bf N}-{\bf 1}-{\bf m}+{\bf j}+{\bf n}}J_{\bf m}^a+2Nh^\vee\lambda_{\bf i}^{{\bf 0};{\bf N}-{\bf 1}}
  \delta_{{\bf j},{\bf 0}}J_{{\bf N}-{\bf 1}}^a
  =\begin{dcases} J_{{\bf i}+{\bf j}}^a,\ &{\bf i}+{\bf j}<{\bf N},\\[.15cm] 0,\ &\mathrm{otherwise}.\end{dcases}
\label{condj}
\ee

For each ${\bf i}$, the conditions (\ref{condj}) for ${\bf j}\neq{\bf 0}$ are all repetitions of conditions appearing for 
${\bf j}={\bf 0}$, so it suffices to consider (\ref{condj}) for ${\bf j}={\bf 0}$:
\be
 2\sum_{{\bf 0}\leq{\bf m}<{\bf N}}\sum_{{\bf 0}\leq{\bf n}\leq{\bf m}}\lambda_{\bf i}^{{\bf n};{\bf N}-{\bf 1}}
  \big(k_{\,{\bf N}-{\bf 1}-{\bf m}+{\bf n}}+Nh^\vee\delta_{{\bf n},{\bf 0}}\delta_{{\bf m},{\bf N}-{\bf 1}}\big)J_{\bf m}^a
  =J_{\bf i}^a.
\label{condj0}
\ee
As the generators $\{J_{\bf m}^a\,|\,{\bf 0}\leq{\bf m}<{\bf N}\}$ are linearly independent,
the constraint (\ref{condj0}) translates into a lower-triangular system of linear equations in the 
variables $\{\lambda_{\bf i}^{{\bf n};{\bf N}-{\bf 1}}\,|\,{\bf 0}\leq{\bf n}<{\bf N}\}$.
Indeed, considering $\lambda_{\bf i}^{\ast;{\bf N}-{\bf 1}}$ as the $N$-vector with components
$\lambda_{\bf i}^{{\bf n};{\bf N}-{\bf 1}}$ ordered canonically according to ${\bf n}\in I_{\bf N}$, such that
\be
 M\lambda_{\bf i}^{\ast;{\bf N}-{\bf 1}}=\begin{pmatrix}\delta_{{\bf j},{\bf i}}\end{pmatrix}_{{\bf 0}\leq{\bf j}<{\bf N}},
 \qquad {\bf 0}\leq{\bf i}<{\bf N},
\label{Mlambda}
\ee
the coefficient matrix $M$ is given by
\be
 M_{{\bf m},{\bf n}}=\begin{dcases} 2k_{{\bf N}-{\bf 1}-{\bf m}+{\bf n}}',\ 
   &{\bf 0}\leq{\bf m}-{\bf n}<{\bf N},\\[.15cm]
   0,\ & \mathrm{otherwise},
  \end{dcases}
\ee
for all ${\bf i}$, where
\be
 k_{\bf m}'=k_{\bf m}+Nh^\vee\delta_{{\bf m},{\bf 0}},\qquad {\bf 0}\leq{\bf m}<{\bf N}.
\label{khvee}
\ee 
All the diagonal entries are thus given by $2k_{{\bf N}-{\bf 1}}$.
The only nonzero component on the righthand side of (\ref{Mlambda}) is a $1$ in the position corresponding 
to ${\bf i}\in I_{\bf N}$.

The structure of $M$ resembles a lower-triangular Toeplitz matrix, but with some entries set to $0$.
Indeed, 
\be
 M=\begin{pmatrix} 
 M_1&&&&\\[.15cm]
 \vdots&\ddots&&&\\
 M_{i_1}&\cdots&M_1&&\\
 \vdots&\ddots&\vdots&\ddots&\\[.1cm]
 M_{N_1}&\cdots&M_{i_1}&\cdots&M_1
 \end{pmatrix},
\label{M}
\ee
where each $M_{i_1}\in\{M_1,\ldots,M_{N_1}\}$ is an $\frac{N}{N_1}\times\frac{N}{N_1}$ lower-triangular
matrix (recall that $N=N_1\ldots N_\sigma$) of the form
\be
 M_{i_1}=\begin{pmatrix} 
 M_{i_1,1}&&&&\\[.15cm]
 \vdots&\ddots&&&\\
 M_{i_1,i_2}&\cdots&M_{i_1,1}&&\\
 \vdots&\ddots&\vdots&\ddots&\\[.1cm]
 M_{i_1,N_2}&\cdots&M_{i_1,i_2}&\cdots&M_{i_1,1}
 \end{pmatrix},
\label{Mi1}
\ee
where each $M_{i_1,i_2}\in\{M_{i_1,1},\ldots,M_{i_1,N_2}\}$ is an $\frac{N}{N_1N_2}\times\frac{N}{N_1N_2}$ 
lower-triangular matrix of similar form, and so on. The innermost lower-triangular matrices appearing in
this nested description of $M$ are $N_\sigma\times N_\sigma$ Toeplitz matrices of the form
\be
 M_{i_1,\ldots,i_{\sigma-1}}=\begin{pmatrix} 
 2k_{i_1,\ldots,i_{\sigma-1},N_\sigma-1}&&&&\\[.15cm]
 \vdots&\ddots&&&\\
 2k_{i_1,\ldots,i_{\sigma-1},N_\sigma-i_\sigma}&&&&\\
 \vdots&\ddots&&\ddots&\\[.1cm]
 2k'_{i_1,\ldots,i_{\sigma-1},0}&\cdots&&\cdots&2k_{i_1,\ldots,i_{\sigma-1},N_\sigma-1}
 \end{pmatrix}.
\label{Msigma}
\ee
For the sequence ${\bf N}=3,2,3$, for example, we thus have
\be
 M=2{\tiny \left(\!\!\begin{array}{cccccccccccccccccc}
  k_{212}&\z&\z&\z&\z&\z&\z&\z&\z&\z&\z&\z&\z&\z&\z&\z&\z&\z\\
  k_{211}&k_{212}&\z&\z&\z&\z&\z&\z&\z&\z&\z&\z&\z&\z&\z&\z&\z&\z\\
  k_{210}&k_{211}&k_{212}&\z&\z&\z&\z&\z&\z&\z&\z&\z&\z&\z&\z&\z&\z&\z\\
  k_{202}&\z&\z&k_{212}&\z&\z&\z&\z&\z&\z&\z&\z&\z&\z&\z&\z&\z&\z\\
  k_{201}&k_{202}&\z&k_{211}&k_{212}&\z&\z&\z&\z&\z&\z&\z&\z&\z&\z&\z&\z&\z\\
  k_{200}&k_{201}&k_{202}&k_{210}&k_{211}&k_{212}&\z&\z&\z&\z&\z&\z&\z&\z&\z&\z&\z&\z\\
  k_{112}&\z&\z&\z&\z&\z&k_{212}&\z&\z&\z&\z&\z&\z&\z&\z&\z&\z&\z\\
  k_{111}&k_{112}&\z&\z&\z&\z&k_{211}&k_{212}&\z&\z&\z&\z&\z&\z&\z&\z&\z&\z\\
  k_{110}&k_{111}&k_{112}&\z&\z&\z&k_{210}&k_{211}&k_{212}&\z&\z&\z&\z&\z&\z&\z&\z&\z\\
  k_{102}&\z&\z&k_{112}&\z&\z&k_{202}&\z&\z&k_{212}&\z&\z&\z&\z&\z&\z&\z&\z\\
  k_{101}&k_{102}&\z&k_{111}&k_{112}&\z&k_{201}&k_{202}&\z&k_{211}&k_{212}&\z&\z&\z&\z&\z&\z&\z\\
  k_{100}&k_{101}&k_{102}&k_{110}&k_{111}&k_{112}&k_{200}&k_{201}&k_{202}&k_{210}&k_{211}&k_{212}&\z&\z&\z&\z&\z&\z\\
  k_{012}&\z&\z&\z&\z&\z&k_{112}&\z&\z&\z&\z&\z&k_{212}&\z&\z&\z&\z&\z\\
  k_{011}&k_{012}&\z&\z&\z&\z&k_{111}&k_{112}&\z&\z&\z&\z&k_{211}&k_{212}&\z&\z&\z&\z\\
  k_{010}&k_{011}&k_{012}&\z&\z&\z&k_{110}&k_{111}&k_{112}&\z&\z&\z&k_{210}&k_{211}&k_{212}&\z&\z&\z\\
  k_{002}&\z&\z&k_{012}&\z&\z&k_{102}&\z&\z&k_{112}&\z&\z&k_{202}&\z&\z&k_{212}&\z&\z\\
  k_{001}&k_{002}&\z&k_{011}&k_{012}&\z&k_{101}&k_{102}&\z&k_{111}&k_{112}&\z&k_{201}&k_{202}&\z&k_{211}&k_{212}&\z\\
  k'_{000}&k_{001}&k_{002}&k_{010}&k_{011}&k_{012}&k_{100}&k_{101}&k_{102}&k_{110}&k_{111}&k_{112}&k_{200}&k_{201}&k_{202}&k_{210}&k_{211}&k_{212}\\
 \end{array}\!\!\right)},
\ee
written using the simplified notation $k_{i_1i_2i_3}=k_{i_1,i_2,i_3}$.

The inverse of $M$,
\be
 M^{-1}=\begin{pmatrix}b_{{\bf m},{\bf n}}\end{pmatrix}_{{\bf 0}\leq{\bf m},{\bf n}<{\bf N}},
\label{Minv}
\ee
has the same nested Toeplitz-like structure, with all diagonal entries given by $1/(2k_{{\bf N}-{\bf 1}})$.
With this, we solve (\ref{Mlambda}) and find
\be
 \lambda_{\bf i}^{{\bf n};{\bf N}-{\bf 1}}=b_{{\bf n},{\bf i}}.
\ee
The Galilean Sugawara construction (\ref{Ti}) is thus given by
\be
 T_{\bf i}=\sum_{{\bf i}\leq{\bf n}<{\bf N}}b_{{\bf n},{\bf i}}
  \sum_{{\bf 0}\leq{\bf t}<{\bf N}-{\bf n}}
   \kappa_{ab}(J_{{\bf n}+{\bf t}}^aJ_{{\bf N}-{\bf 1}-{\bf t}}^b).
\label{Tifinal}
\ee

For each ${\bf i}$, the value of the central parameter $c_{\bf i}$ follows from the leading pole in the OPE
\be
  T_{\bf 0}(z)T_{\bf i}(w)\sim\sum_{{\bf i}\leq{\bf n}<{\bf N}}b_{{\bf n},{\bf i}}
  \sum_{{\bf 0}\leq{\bf t}<{\bf N}-{\bf n}}
  \frac{\kappa_{ab}\kappa^{ab}k_{{\bf N}-{\bf 1}+{\bf n}}}{(z-w)^4}
  +\frac{2T_{\bf i}(w)}{(z-w)^2}+\frac{\pa T_{\bf i}(w)}{z-w}.
\ee
Since $k_{\bf h} = 0$ unless ${\bf h}<{\bf N}$, the only contribution to the leading-pole term 
appears for ${\bf n}={\bf 0}$, hence for ${\bf i}={\bf 0}$. The term thus reduces to
\be
  \frac{2\,\delta_{{\bf i},{\bf 0}}}{k_{{\bf N}-{\bf 1}}}
  \sum_{{\bf 0}\leq{\bf t}<{\bf N}}
  \frac{\kappa_{ab}\kappa^{ab}k_{{\bf N}-{\bf 1}}}{(z-w)^4}
  =\frac{2N\dim\g\,\delta_{{\bf i},{\bf 0}}}{(z-w)^4},
\ee
from which it follows that
\be
 c_{\bf 0}=N\dim\g,\qquad c_{\bf i}=0,\qquad {\bf i}\neq{\bf 0}.
\label{cSug}
\ee
This result for the central charge $c_{\bf 0}$ resembles similar results \cite{FOFS94}
for Sugawara constructions associated with so-called double extensions \cite{MR85}.

\paragraph{$\Vir_G^{2,3}$ algebra:}
To illustrate, we consider the contraction sequence ${\bf N}=2,3$. In this case, $\sigma=2$ and $N=6$, 
while the canonical ordering is 
\be
 I_{2,3}:\ 0,0;\ 0,1;\ 0,2;\ 1,0;\ 1,1;\ 1,2.
\ee
The corresponding Galilean Sugawara construction is given by
\begin{align}
 T_{0,0}&=\frac{\kappa_{ab}}{2k_{1,2}}\big[
  (J_{0,0}^aJ_{1,2}^b)+(J_{0,1}^aJ_{1,1}^b)+(J_{0,2}^aJ_{1,0}^b)+(J_{1,0}^aJ_{0,2}^b)+(J_{1,1}^aJ_{0,1}^b)
   +(J_{1,2}^aJ_{0,0}^b)
\nonumber\\[.15cm]
 &-\tfrac{k_{1,1}}{k_{1,2}}\big((J_{0,1}^aJ_{1,2}^b)+(J_{0,2}^aJ_{1,1}^b)+(J_{1,1}^aJ_{0,2}^b)
   +(J_{1,2}^aJ_{0,1}^b)\big)
  +\tfrac{(k_{1,1})^2-k_{1,0}k_{1,2}}{(k_{1,2})^2}\big((J_{0,2}^aJ_{1,2}^b)+(J_{1,2}^aJ_{0,2}^b)\big)
\nonumber\\[.15cm]
 &-\tfrac{k_{0,2}}{k_{1,2}}\big((J_{1,0}^aJ_{1,2}^b)+(J_{1,1}^aJ_{1,1}^b)+(J_{1,2}^aJ_{1,0}^b)
   +\tfrac{2k_{0,2}k_{1,1}-k_{0,1}k_{1,2}}{(k_{1,2})^2}\big((J_{1,1}^aJ_{1,2}^b)+(J_{1,2}^aJ_{1,1}^b)\big)
\nonumber\\[.15cm]
 &-\tfrac{3k_{0,2}(k_{1,1})^2-2(k_{0,2}k_{1,0}+k_{0,1}k_{1,1})k_{1,2}+k_{0,0}'(k_{1,2})^2}{(k_{1,2})^3}
   (J_{1,2}^aJ_{1,2}^b)  \big],
\\[.15cm]
 T_{0,1}&=\frac{\kappa_{ab}}{2k_{1,2}}\big[
 (J_{0,1}^aJ_{1,2}^b)+(J_{0,2}^aJ_{1,1}^b)+(J_{1,1}^aJ_{0,2}^b)+(J_{1,2}^aJ_{0,1}^b)
  -\tfrac{k_{1,1}}{k_{1,2}}\big((J_{0,2}^aJ_{1,2}^b)+(J_{1,2}^aJ_{0,2}^b)\big)
\nonumber\\[.15cm]
 &-\tfrac{k_{0,2}}{k_{1,2}}\big((J_{1,1}^aJ_{1,2}^b)+(J_{1,2}^aJ_{1,1}^b)\big)
  +\tfrac{2k_{0,2}k_{1,1}-k_{0,1}k_{1,2}}{(k_{1,2})^2}(J_{1,2}^aJ_{1,2}^b)\big],
\\[.15cm]
 T_{0,2}&=\frac{\kappa_{ab}}{2k_{1,2}}\big[
  (J_{0,2}^aJ_{1,2}^b)+(J_{1,2}^aJ_{0,2}^b)-\tfrac{k_{0,2}}{k_{1,2}}(J_{1,2}^aJ_{1,2}^b)\big],
\\[.15cm]
 T_{1,0}&=\frac{\kappa_{ab}}{2k_{1,2}}\big[
  (J_{1,0}^aJ_{1,2}^b)+(J_{1,1}^aJ_{1,1}^b)+(J_{1,2}^aJ_{1,0}^b)
  -\tfrac{k_{1,1}}{k_{1,2}}\big((J_{1,1}^aJ_{1,2}^b)+(J_{1,2}^aJ_{1,1}^b)\big)
  +\tfrac{(k_{1,1})^2-k_{1,0}k_{1,2}}{(k_{1,2})^2}(J_{1,2}^aJ_{1,2}^b)\big],
\\[.15cm]
 T_{1,1}&=\frac{\kappa_{ab}}{2k_{1,2}}\big[
  (J_{1,1}^aJ_{1,2}^b)+(J_{1,2}^aJ_{1,1}^b)-\tfrac{k_{1,1}}{k_{1,2}}(J_{1,2}^aJ_{1,2}^b)\big],
\\[.15cm]
 T_{1,2}&=\frac{\kappa_{ab}}{2k_{1,2}}(J_{1,2}^aJ_{1,2}^b),
\end{align}
and has central parameters
\be
  c_{0,0}=6\dim\g,\qquad c_{0,1}=c_{0,2}=c_{1,0}=c_{1,1}=c_{1,2}=0.
\label{c00}
\ee

%%%%%%%%%%%%%%%%%%%%%%%%%%
\subsection{Sugawara before Galilean contraction}
\label{Sec:SugCon}
%%%%%%%%%%%%%%%%%%%%%%%%%%

As above, let ${\bf N}=N_1,\ldots,N_\sigma$ and $N=N_1\ldots N_\sigma$.
Accordingly, on the individual factors of $\gh^{\,\otimes N}$\!, we denote the Sugawara construction by
\be
 T_{({\bf i})}=\frac{\kappa_{ab}}{2(k_{({\bf i})}+h^\vee)}(J_{({\bf i})}^aJ_{({\bf i})}^b),\qquad
 c_{({\bf i})}=\frac{k_{({\bf i})}\dim\g}{k_{({\bf i})}+h^\vee},\qquad {\bf 0}\leq{\bf i}<{\bf N},
\ee
let ${\bf T}_{*}$ and ${\bf c}_{*}$ denote the corresponding $N$-vectors formed as in (\ref{Astar}), 
and change basis as in (\ref{AAA}):
\be
 {\bf T}_{\eps}
  =\begin{pmatrix} T_{{\bf i},\eps}\end{pmatrix}_{{\bf 0}\leq{\bf i}<{\bf N}}=U_{\bf N}(\eb,\ob){\bf T}_{*},\qquad
 {\bf c}_{\eps}=\begin{pmatrix} c_{{\bf i},\eps}\end{pmatrix}_{{\bf 0}\leq{\bf i}<{\bf N}}
  =U_{\bf N}(\eb,\ob){\bf c}_{*}.
\label{Teps}
\ee
It follows that
\be
 T_{{\bf i},\eps}=\frac{\displaystyle{\sum_{{\bf 0}\leq{\bf j},{\bf n},{\bf n}'<{\bf N}} %\sum_{a,b=1}^{\dim\g}
  \Big(\prod_{\ell=1}^\sigma(\eps_\ell\omega_\ell^{j_\ell})^{N_\ell-1+i_\ell-n_\ell-n_\ell'}\Big)
    \kappa_{ab}(J_{{\bf n},\eps}^aJ_{{\bf n}',\eps}^b)}}{\displaystyle{2Nk_{{\bf N}-{\bf 1},\eps}
     \sum_{{\bf 0}\leq{\bf m}<{\bf N}}a_{{\bf m}}
   \prod_{\ell=1}^\sigma(\eps_\ell\omega_\ell^{j_\ell})^{m_\ell}}},
\ee
where
\be
 a_{{\bf m}}=\frac{k_{{\bf N}-{\bf 1}-{\bf m},\eps}+Nh^\vee\delta_{{\bf m},{\bf N}-{\bf 1}}}{k_{{\bf N}-{\bf 1},\eps}},
 \qquad {\bf 0}\leq{\bf m}<{\bf N}.
\label{am}
\ee
Since $a_{{\bf 0}}=1$, there exist $\hat{a}_{{\bf m}}$, ${\bf 0}\leq{\bf m}<{\bf N}$, 
where $\hat{a}_{{\bf 0}}=1$, such that
\be
 \Big(\sum_{{\bf 0}\leq{\bf m}<{\bf N}}a_{{\bf m}}
  \prod_{\ell=1}^\sigma(\eps_\ell\omega_\ell^{j_\ell})^{m_\ell}\Big)^{-1}
 =\sum_{{\bf 0}\leq{\bf m}<{\bf N}}\hat{a}_{{\bf m}}
  \prod_{\ell=1}^\sigma(\eps_\ell\omega_\ell^{j_\ell})^{m_\ell}
  +\Oc(\eps_1^{N_1},\ldots,\eps_\sigma^{N_\sigma}).
\ee
We can thus write
\be
 T_{{\bf i},\eps}=\frac{1}{2Nk_{{\bf N}-{\bf 1},\eps}}\sum_{{\bf 0}\leq{\bf j},{\bf n},{\bf n}',{\bf m}<{\bf N}}
  \hat{a}_{{\bf m}}
  \Big(\prod_{\ell=1}^\sigma(\eps_\ell\omega_\ell^{j_\ell})^{N_\ell-1+i_\ell-n_\ell-n_\ell'+m_\ell}\Big)
    \kappa_{ab}(J_{{\bf n},\eps}^aJ_{{\bf n}',\eps}^b)
  +\Oc(\eps_1^{N_1},\ldots,\eps_\sigma^{N_\sigma}),
\ee
For each $\ell\in\{1,\ldots,\sigma\}$, the summation over $j_\ell$ yields a factor of the form
\be
 \sum_{j_\ell=0}^{N_\ell-1}\omega_\ell^{j_\ell(N_\ell-1+i_\ell-n_\ell-n_\ell'+m_\ell)}
  =\begin{dcases} N_\ell,\ & N_\ell-1+i_\ell-n_\ell-n_\ell'+m_\ell\equiv 0\ \ 
  (\mathrm{mod}\ N_\ell),\\[.15cm]
  0,\ & N_\ell-1+i_\ell-n_\ell-n_\ell'+m_\ell\not\equiv 0\ \ (\mathrm{mod}\ N_\ell), \end{dcases}
\ee
so
\be
 \sum_{{\bf 0}\leq{\bf j}<{\bf N}}
  \prod_{\ell=1}^\sigma \omega_\ell^{j_\ell(N_\ell-1+i_\ell-n_\ell-n_\ell'+m_\ell)}
 =\begin{dcases} N,\ & {\bf N}-{\bf 1}+{\bf i}-{\bf n}-{\bf n}'+{\bf m}\equiv {\bf 0}\ \ 
  (\mathrm{mod}\ {\bf N}),\\[.15cm]
  0,\ & {\bf N}-{\bf 1}+{\bf i}-{\bf n}-{\bf n}'+{\bf m}\not\equiv {\bf 0}\ \ (\mathrm{mod}\ {\bf N}). \end{dcases}
\ee 
Since ${\bf N}-{\bf 1}+{\bf i}-{\bf n}-{\bf n}'+{\bf m}>-{\bf N}$, it follows that the $T_{{\bf i},\eps}$-coefficients
to negative powers of any of the $\eps_\ell$'s are all zero.
The limit $\eb\to{\bf 0}$ is therefore well-defined, and we find
\begin{align}
 T_{{\bf i},\eps}\to T_{\bf i}&=\frac{1}{2k_{{\bf N}-{\bf 1}}}\sum_{{\bf 0}\leq{\bf n},{\bf n}',{\bf m}<{\bf N}}
  \hat{a}_{\bf m}
    \kappa_{ab}(J_{{\bf n}}^aJ_{{\bf n}'}^b)\delta_{{\bf N}-{\bf 1}+{\bf i}-{\bf n}-{\bf n}'+{\bf m},{\bf 0}}
 \nonumber\\[.15cm]
 &=\sum_{{\bf i}\leq{\bf n}<{\bf N}}\frac{\hat{a}_{{\bf n}-{\bf i}}}{2k_{{\bf N}-{\bf 1}}}
 \sum_{{\bf 0}\leq{\bf t}<{\bf N}-{\bf n}} 
    \kappa_{ab}(J_{{\bf n}+{\bf t}}^aJ_{{\bf N}-{\bf 1}-{\bf t}}^b).
\end{align}
This is seen to agree with (\ref{Tifinal}) if
\be
 b_{{\bf n},{\bf i}}=\frac{\hat{a}_{{\bf n}-{\bf i}}}{2k_{{\bf N}-{\bf 1}}}
\label{ba}
\ee
for all ${\bf i}\leq{\bf n}<{\bf N}$, that is, if
\be
  \lambda_{\bf i}^{{\bf n};{\bf N}-{\bf 1}}=\frac{\hat{a}_{{\bf n}-{\bf i}}}{2k_{{\bf N}-{\bf 1}}}.
\label{laa}
\ee
In the affirmative, the relations (\ref{laa}) follow from the similarity in structures of $M_{{\bf m},{\bf n}}$ in
(\ref{Mlambda}) and $a_{\bf m}$ in (\ref{am}).
We have thus established the commutativity of Galilean contractions and Sugawara constructions, 
{\em without} explicit knowledge of the
coefficients $b_{{\bf n},{\bf i}}$ and $\hat{a}_{\bf n}$ appearing in the inversion of the coefficient
matrix $M$ and the series expansion of the denominator of $T_{{\bf i},\eps}$, respectively.
The coefficients are readily obtained case by case, but cumbersome to express for general parameters.
For $\sigma=1$, the coefficients are given in \cite{RR19}.

For the central parameters, we have
\begin{align}
 c_{{\bf i},\eps}&=\frac{\displaystyle{\sum_{{\bf 0}\leq{\bf j},{\bf n}<{\bf N}}
  \Big(\prod_{\ell=1}^\sigma(\eps_\ell\omega_\ell^{j_\ell})^{N_\ell-1+i_\ell-n_\ell}\Big)
    k_{{\bf n},\eps}}\dim\g}{\displaystyle{k_{{\bf N}-{\bf 1},\eps}
     \sum_{{\bf 0}\leq{\bf m}<{\bf N}}a_{{\bf m}}
   \prod_{\ell=1}^\sigma(\eps_\ell\omega_\ell^{j_\ell})^{m_\ell}}}
 \nonumber\\[.15cm]
 &=\frac{\dim\g}{k_{{\bf N}-{\bf 1},\eps}}\sum_{{\bf 0}\leq{\bf j},{\bf n},{\bf m}<{\bf N}}\hat{a}_{\bf m}k_{{\bf n},\eps}
   \prod_{\ell=1}^\sigma(\eps_\ell\omega_\ell^{j_\ell})^{N_\ell-1+i_\ell-n_\ell+m_\ell}
   +\Oc(\eps_1^{N_1},\ldots,\eps_\sigma^{N_\sigma}),
\end{align}
so in the limit $\eb\to{\bf 0}$,
\be
 c_{{\bf i},\eps}\to c_{\bf i}
  =\frac{N\dim\g}{k_{{\bf N}-{\bf 1}}}
    \sum_{{\bf 0}\leq{\bf n},{\bf m}<{\bf N}}\hat{a}_{\bf m}k_{{\bf n}}
    \delta_{{\bf N}-{\bf 1}+{\bf i}-{\bf n}+{\bf m},{\bf 0}}
  =N\dim\g\,\delta_{{\bf i},{\bf 0}},
\ee
in accordance with (\ref{cSug}).

%%%%%%%%%%%%%%%%%%%%%%%%%%
\section{Galilean $W_3$ algebras}
\label{Sec:GW3}
%%%%%%%%%%%%%%%%%%%%%%%%%%

Our generalised Galilean contractions can also be applied to W-algebras. Below, we present the results for 
the factorisation sequence ${\bf N}=2,3$ applied to the $W_{3}$ algebra, giving rise to the sixth-order Galilean 
algebra $(W_3)_G^{2,3}$. In preparation, we first recall the structure of the $W_3$ algebra and its second- 
and third-order Galilean counterparts.

%%%%%%%%%%%%%%%%%%%%%%%%%%
\subsection{$W_3$ algebra}
\label{Sec:W3}
%%%%%%%%%%%%%%%%%%%%%%%%%%

The $W_3$ algebra \cite{Zam85} of central charge $c$
is generated by a Virasoro field $T$ and a primary field $W$ of conformal weight $3$, with star relations
\be
 T\ast T\simeq\tfrac{c}{2}\{\id\}+2\{T\},\qquad T\ast W\simeq3\{W\},\qquad 
 W\ast W\simeq\tfrac{c}{3}\{\id\}+2\{T\}+\tfrac{32}{22+5c}\{\La\},
\ee
where
\be
 \La=(TT)-\tfrac{3}{10}\pa^2T
\ee
is quasi-primary.

%%%%%%%%%%%%%%%%%%%%%%%%%%
\subsection{Galilean $W_3$ algebras of type $(W_3)^N_G$}
\label{Sec:W3N2N3}
%%%%%%%%%%%%%%%%%%%%%%%%%%

Following \cite{RR19} and Section \ref{Sec:Higher}, the $N$th-order Galilean algebra $(W_3)_G^N$ is generated 
by the fields $\{T_i,W_i\,|\,i=0,\ldots,N-1\}$ and has central parameters $\{c_i\,|\,i=0,\ldots,N-1\}$. 
The star relations involving the $T_i$ fields are
\be
 T_i\ast T_j\simeq\tfrac{c_{i+j}}{2}\{\id\}+2\{T_{i+j}\},\qquad 
 T_i\ast W_j\simeq 3\{W_{i+j}\},\qquad i+j\in\{0,\ldots,N-1\},
\label{QPrels}
\ee
and
\be
 T_i\ast T_j\simeq T_i\ast W_j\simeq0,\qquad i+j\geq N.
\label{TTT}
\ee
As will become clear below, it is convenient to introduce
\be
 c_0'=c_0+\tfrac{22N}{5}.
\label{cop}
\ee

\paragraph{$(W_3)^2_G$ algebra:}
For $N=2$, the star relations between the $W$-fields are given by
\begin{align}
 W_0\ast W_0&\simeq\tfrac{c_{0}}{3}\{\id\}+2\{T_0\}
  +\tfrac{64}{5c_1}\{\La_{0,1}\}-\tfrac{32c_0'}{5c_1^2}\{\La_{1,1}\},
 \\[.15cm]
  W_0\ast W_1&\simeq \tfrac{c_1}{3}\{\id\}+2\{T_1\}+\tfrac{32}{5c_1}\{\La_{1,1}\},
 \\[.15cm]
  W_1\ast W_1&\simeq0,
\end{align}
where 
\be
 \La_{0,1}=(T_0T_1)-\tfrac{3}{10}\pa^{2}T_{1},\qquad \La_{1,1}=(T_1T_1)
\ee
are quasi-primary with respect to $T_0$.

\paragraph{$(W_3)^3_G$ algebra:}
For $N=3$, the star relations between the $W$-fields are given by
\begin{align}
 W_0\ast W_0&\simeq
  \tfrac{c_0}{3}\{\id\}
  +2\{T_0\}
  +\tfrac{64}{5c_2}\{\La_{0,2}\}
  +\tfrac{32}{5c_2}\{\La_{1,1}\}
  -\tfrac{64c_1}{5(c_2)^2}\{\La_{1,2}\}
  +\tfrac{32[(c_1)^2-c_0'c_2]}{5(c_2)^3}\{\La_{2,2}\},
\\[.15cm]
 W_0\ast W_1&\simeq
  \tfrac{c_1}{3}\{\id\}
  +2\{T_1\}
  +\tfrac{64}{5c_2}\{\La_{1,2}\}
  -\tfrac{32c_1}{5(c_2)^2}\{\La_{2,2}\},
\\[.15cm]
 W_0\ast W_2&\simeq
  \tfrac{c_2}{3}\{\id\}
  +2\{T_2\}
  +\tfrac{32}{5c_2}\{\La_{2,2}\},
\end{align}
and
\be
 W_1\ast W_1\simeq W_0\ast W_2,\qquad
 W_1\ast W_2\simeq W_2\ast W_2\simeq0,
\ee
where
\be
 \La_{0,2}=(T_0T_2)-\tfrac{3}{10}\pa^2T_2,\qquad
 \La_{1,1}=(T_1T_1)-\tfrac{3}{10}\pa^2T_2,\qquad
 \La_{1,2}=(T_1T_2),\qquad
 \La_{2,2}=(T_2T_2)
\ee
are quasi-primary with respect to $T_0$.
Both $(W_3)_G^2$ and $(W_3)_G^3$ are readily seen to be (multi-)graded, truncated according to their order
($2$ and $3$, respectively).

%%%%%%%%%%%%%%%%%%%%%%%%%%
\subsection{Galilean algebra $(W_3)^{2,3}_G$}
\label{Sec:W3N23}
%%%%%%%%%%%%%%%%%%%%%%%%%%

The $N$th-order Galilean algebra $(W_3)_G^{\bf N}$ is generated by the fields 
$\{T_{\bf i},W_{\bf i}\,|\,{\bf 0}\leq{\bf i}<{\bf N}\}$ and has central parameters 
$\{c_{\bf i}\,|\,{\bf 0}\leq{\bf i}<{\bf N}\}$, with the star relations involving the $T$ fields given by (\ref{TiTj}) and
\be
 T_{\bf i}\ast W_{\bf j}\simeq 
 \begin{dcases}3\{W_{{\bf i}+{\bf j}}\}, \ &{\bf i}+{\bf j}<{\bf N}, \\[.15cm] 
 0, \ &\mathrm{otherwise}.
 \end{dcases}
\label{TW3W}
\ee
Extending (\ref{cop}), for all ${\bf N}$, it is convenient to introduce
\be
 c_{\bf 0}'=c_{\bf 0}+\tfrac{22N}{5}.
\ee
Comparing this with (\ref{khvee}), one may view $h^\vee=\frac{22}{5}$ as the associated
``dual Coxeter number".

\paragraph{$(W_3)^{2,3}_G$ algebra:}
For the contraction sequence ${\bf N}=2,3$, up to the equivalences (\ref{ijij}), 
the inequivalent nontrivial star relations involving the $W$ fields are given by
\begin{align}
 W_{0,0}\ast W_{0,0}&\simeq\tfrac{c_{0,0}}{3}\{\id\}+2\{T_{0,0}\}
  +\tfrac{64}{5c_{1,2}}\{\La_{0,0;1,2}+\La_{0,1;1,1}+\La_{0,2;1,0}\}
  -\tfrac{64c_{1,1}}{5(c_{1,2})^2}\{\La_{0,1;1,2}+\La_{0,2;1,1}\}
\nonumber\\[.15cm]
 &+\tfrac{64[(c_{1,1})^2-c_{1,0}c_{1,2}]}{5(c_{1,2})^3}\{\La_{0,2;1,2}\}
  -\tfrac{32c_{0,2}}{5(c_{1,2})^2}\{2\La_{1,0;1,2}+\La_{1,1;1,1}\}
  +\tfrac{64[2c_{0,2}c_{1,1}-c_{0,1}c_{1,2}]}{5(c_{1,2})^3}\{\La_{1,1;1,2}
\}\nonumber\\[.15cm]
 &-\tfrac{32[3c_{0,2}(c_{1,1})^2-2(c_{0,1}c_{1,1}+c_{0,2}c_{1,0})c_{1,2}+c_{0,0}'(c_{1,2})^2]}{5(c_{1,2})^4}\{\La_{1,2;1,2}\},
\\[.15cm]
 W_{0,0}\ast W_{0,1}&\simeq\tfrac{c_{0,1}}{3}\{\id\}+2\{T_{0,1}\}
  +\tfrac{64}{5c_{1,2}}\{\La_{0,1;1,2}+\La_{0,2;1,1}\}-\tfrac{64c_{1,1}}{5(c_{1,2})^2}\{\La_{0,2;1,2}\}
  -\tfrac{64c_{0,2}}{5(c_{1,2})^2}\{\La_{1,1;1,2}\}
\nonumber\\[.15cm]
 &+\tfrac{32[2c_{0,2}c_{1,1}-c_{0,1}c_{1,2}]}{5(c_{1,2})^3}\{\La_{1,2;1,2}\},
\\[.15cm]
 W_{0,0}\ast W_{0,2}&\simeq\tfrac{c_{0,2}}{3}\{\id\}+2\{T_{0,2}\}+\tfrac{64}{5c_{1,2}}\{\La_{0,2;1,2}\}
  -\tfrac{32c_{0,2}}{5(c_{1,2})^2}\{\La_{1,2;1,2}\},
\\[.15cm]
 W_{0,0}\ast W_{1,0}&\simeq\tfrac{c_{1,0}}{3}\{\id\}+2\{T_{1,0}\}
  +\tfrac{32}{5c_{1,2}}\{2\La_{1,0;1,2}+\La_{1,1;1,1}\}-\tfrac{64c_{1,1}}{5(c_{1,2})^2}\{\La_{1,1;1,2}\}
\nonumber\\[.15cm]
 &+\tfrac{32[(c_{1,1})^2-c_{1,0}c_{1,2}]}{5(c_{1,2})^3}\{\La_{1,2;1,2}\},
\\[.15cm]
 W_{0,0}\ast W_{1,1}&\simeq\tfrac{c_{1,1}}{3}\{\id\}+2\{T_{1,1}\}+\tfrac{64}{5c_{1,2}}\{\La_{1,1;1,2}\}
  -\tfrac{32c_{1,1}}{5(c_{1,2})^2}\{\La_{1,2;1,2}\},
\\[.15cm]
 W_{0,0}\ast W_{1,2}&\simeq\tfrac{c_{1,2}}{3}\{\id\}+2\{T_{1,2}\}+\tfrac{32}{5c_{1,2}}\{\La_{1,2;1,2}\},
\end{align}
where, for ${\bf i}+{\bf j}\geq{\bf N}-{\bf 1}$,
\be
 \La_{{\bf i};\,{\bf j}}=(T_{\bf i}\,T_{\bf j})-\tfrac{3}{10}\pa^2T_{{\bf N}-{\bf 1}}\delta_{{\bf i}+{\bf j},\,{\bf N}-{\bf 1}}
\label{Laij}
\ee
is quasi-primary with respect to the Virasoro generator $T_{\bf 0}$. It follows that the sixth-order Galilean algebra 
$(W_3)_G^{2,3}$ is multi-graded, with truncation dictated by the sequence $2,3$.

%%%%%%%%%%%%%%%%%%%%%%%%%%
\section{Discussion}
\label{Sec:Discussion}
%%%%%%%%%%%%%%%%%%%%%%%%%%

In our continued exploration \cite{ChrisThesis,RR17,RR19} of Galilean contractions, we have presented a 
generalisation of the contraction procedure to multi-graded Galilean algebras. 
Our construction uses factorisations of the order parameter $N$, and has resulted in whole new families
of higher-order Galilean conformal algebras, including Virasoro, affine Kac-Moody and $W_{3}$ algebras.
We have also discussed how some of these algebras are related to a multivariable extension of Takiff algebras.

$W$-algebras related to Takiff algebras were introduced in \cite{MR19,Mol20} and
constructed as principal $W$-algebras built on the centralizer of a nilpotent element
in $\mathfrak{gl}(n)$. The construction is carried out in the context of (Poisson) vertex algebras,
and it appears natural that it is linked to the one presented here. In particular, the nilpotent
element being characterised by a partition $\lambda=\lambda_1,\lambda_2,\ldots$, the algebras have
an indexation comparable to ${\bf N}=N_1,N_2,\ldots$ used in our multi-graded contraction procedure.

Other avenues for future work include {\em asymmetric contractions} and {\em free-field realisations}.
Asymmetric Galilean $N=1$ superconformal algebras were constructed in 
\cite{BDMT14,BDMT17,CGOR16,BJLMN16} from a Galilean contraction of
the tensor product $\mathfrak{SVir}\otimes\Vir$, where one contracts the Virasoro subalgebra 
of an $N=1$ superconformal algebra, $\mathfrak{SVir}$, 
with a separate Virasoro algebra. The ensuing Galilean superconformal algebra can be viewed as encoding 
a $(1,0)$ supersymmetry. This was extended in \cite{RR19} to a contraction of the asymmetric tensor 
product $W_3\otimes\Vir$, giving rise to a Galilean $W_3$ algebra generated by fields $T_0, T_1, W$.
There is significant freedom in such contractions, and we hope to return elsewhere with a partial
classification of the inequivalent Galilean algebras that can arise this way.

Free-field realisations \cite{DF84,Wak86,FMS86,FL88,FF90,BS93,Fre94,Ras96,Ras98,Kau00} 
are ubiquitous in conformal field theory, and we find it natural to expect that
they will continue to play a central role when Galilean conformal symmetries are present.
Some work on this has been done \cite{BJMN16,AR16,BJLMN16}, 
but a systematic approach and general results remain outstanding. We hope to report such advances
in the near future.

%%%%%%%%%%%%%%%%%%%%%%%%%%
\subsection*{Acknowledgements}
%%%%%%%%%%%%%%%%%%%%%%%%%%

ER thanks The SMRI International Visitor program at The University of Sydney, and 
The University of Queensland (UQ), for their support during his visit to UQ in September 2019.
JR was supported by the Australian Research Council under the Discovery Project scheme, 
project number DP160101376. 
CR was funded by a University of Queensland Research Scholarship.

%%%%%%%%%%%%%%%%%%%%%
%

\end{document}